\documentclass[12pt, a4]{article}
\usepackage[utf8]{inputenc}
\usepackage{geometry}
\usepackage{float}
\usepackage[numbers]{natbib}

\usepackage{authblk}
\usepackage{graphicx}
\usepackage{pgfplots}
\usepackage{booktabs} 
\usepackage{multirow}
\usepackage{multicol}
\usepackage{bm}
\usepackage{bbm}
\usepackage{amssymb}
\usepackage{amsbsy}
\usepackage{array}
\usepackage{cite}

\usepackage{microtype}
\usepackage{subfigure}
\usepackage{amsmath}
\usepackage{amsfonts}
\usepackage{mathrsfs}
\usepackage{color}

\title{CNN-based Realized Covariance Matrix Forecasting}

\author{
  Yanwen Fang$^{1}$,~~
  Philip L. H. Yu$^{1,2}$\thanks{Corresponding author. Email address: plhyu@eduhk.hk.},~~
  Yaohua Tang\textsuperscript{1} \\
$^{1}${\footnotesize{}Department of Statistics and Actuarial Science,
The University of Hong Kong}\\
$^{2}${\footnotesize{}Department of Mathematics and Information Technology,
Education University of Hong Kong}}
\date{}

\pgfplotsset{compat=1.15}
\begin{document}

\maketitle

\begin{abstract}
It is well known that modeling and forecasting realized covariance matrices of asset returns play a crucial role in the field of finance. The availability of high frequency intraday data enables the modeling of the realized covariance matrices directly. However, most of the models available in the literature depend on strong structural assumptions and they often suffer from the curse of dimensionality. We propose an end-to-end trainable model built on the CNN and Convolutional LSTM (ConvLSTM) which does not require to make any distributional or structural assumption but could handle high-dimensional realized covariance matrices consistently.  The proposed model focuses on local structures
and spatiotemporal correlations. It learns a nonlinear mapping that connect the historical realized covariance matrices to the future one. Our empirical studies on synthetic and real-world datasets demonstrate its excellent forecasting ability compared with several advanced volatility models.
\end{abstract}

\textbf{Keywords:} Deep Learning, Convolutional Neural Network, Convolutional LSTM, Realized Covariance Matrix.

\section{Introduction}
\label{sec:introduction}
Modeling and forecasting covariance matrices of financial asset returns have long been an important problem in asset pricing, portfolio allocation and risk management.
The availability of high-frequency intraday financial data enables us to estimate daily volatilities and co-volatilities of asset returns directly, which leads to the so-called realized covariance (RCOV) matrices. In the field of finance, modeling RCOV matrices  has always been a hot research topic.

In this modeling, the first issue is that the fitted RCOV matrices should be guaranteed to be symmetric and positive definite. A natural choice for this issue is the family of matrix-valued Wishart distributions which automatically generates random positive definite matrices without imposing additional constraints.
One example is the conditional autoregressive Wishart (CAW) model proposed by \citet{golosnoy2012conditional}. Another approach is to transform the RCOV matrices. \citet{bauer2011forecasting} suggested applying matrix logarithm transformation, and \citet{wang2016some} suggested square root transformation to the RCOV matrices.
High-dimensionality is the second issue. For $d$ assets, there are $d(d + 1)/2$ entries in the RCOV matrices, which means the number of parameters needed to model RCOV matrices grows quickly with $d$.

To overcome the difficulty due to the high-dimensionality, structural assumptions are needed in order to estimate the RCOV matrix consistently. Some assumptions of the existing effective models for the RCOV matrices are so restrictive that they cannot model the dynamic of the RCOV matrices properly.  \citet{tao2011large}  proposed the matrix factor analysis (MFA) model which projects high dimensional RCOV matrices to a time series of low dimensional matrices, and then builds a vector autoregressive (VAR) model to the vectorized time series. \citet{KerenShen2020} replaced the VAR model by a diagonal CAW (DCAW) model with diagonal matrices. As the projected dimension is usually low, say 3-10, both VAR and DCAW models may contain too few parameters to determine the complex dependency structure of a high-dimensional RCOV matrix.

With the recent advancement in powerful computing resources (like GPUs \citep{krizhevsky2012imagenet}) and the easy access to an abundance of data (like ImageNet \citep{deng2009imagenet}), deep learning models have received great attention. They are proved to be excellent in solving hard problems, like speech recognition, object recognition, and machine translation where the data are notably complex, and they have won numerous contests in pattern recognition and machine learning, beating  the stat-of-art traditional methods. Convolutional neural network (CNN) and recurrent neural network (RNN) are the most common types of neural networks, which capture image data and sequential data well respectively. Long short-term memory (LSTM) as a special RNN structure has been proven to perform extremely well on temporal data \citep{hochreiter1997long, gers2002learning, sundermeyer2012lstm, gers2002applying, cao2019financial, srivastava2015unsupervised, sutskever2014sequence}. If a sequence of RCOV matrices can be viewed as a sequence of maps, the goal of our task is to give precise prediction of future maps given the previously observed maps, which in essence is a spatiotemporal sequence forecasting problem. \citet{xingjian2015convolutional} extended the traditional LSTM models to have convolutional structures, called ConvLSTM. ConvLSTM performs well on spatiotemporal sequence forecasting problem which provides some novel insights on our task.

In this paper, we take the local structure of RCOV matrices into consideration and formulate RCOV matrices forecasting problem as a spatiotemporal sequence forecasting problem.
Base on this, we propose an end-to-end forecasting model for this problem. The model has almost no assumption on the RCOV matrices and has a great flexibility that the model could fit to both low-dimensional and high-dimensional matrices. Unlike other statistical models, the parameter size of our model does not grow dramatically as the data dimension grows. The proposed CNN and ConvLSTM based model detects local structures from the historical RCOV matrices and try to forecast the future RCOV matrix after applying non-linear mapping to these patterns. We will see later that our proposed model could obtain a better estimation accuracy in much shorter time than the existing methods studied.

The rest of this paper is organized as follows. Section \ref{sec:background} gives a brief review of recent development of time series modeling for RCOV matrices and some preliminary knowledge about CNN and ConvLSTM. Our model structure is proposed in Section \ref{sec:ourmodel}. Simulation results on synthetic data and  experimental results on real data are reported in Section \ref{sec:emprical}.
Section \ref{sec:conclusion} concludes with some final remarks and discusses some future work.

\section{Preliminaries}
\label{sec:background}
\subsection{Realized Covariance Matrix Modeling}
Suppose that there are $d$ assets and their log price process $\bm{X}_t = \left\{ X_1(t),\dots X_d(t) \right\}^\top$ obeys a continuous diffusion model
\begin{equation}
d\bm{X}(t) = \bm{\mu}_tdt + \bm{\sigma}_td\bm{W}_t, t\in[0,L],
\end{equation}
where $L$ is an integer, $\bm{W}_t$ is a $d$-dimensional standard Brownian motion, $\bm{\mu}_t$ is a drift taking values in $\mathbb{R}^d$, and $\bm{\sigma}_t$ is a $d\times d$ matrix.  $\bm{\mu}_t$ and $\bm{\sigma}_t$ are assumed to be continuous in $t$. Let a day be a unit time. The integrated volatility matrix for the $t$-th day is defined  as
\begin{equation}
\bm{\Sigma}_x(t) = \int_{t-1}^{t}\bm{\sigma}_s \bm{\sigma}_s^{\top}ds, ~~~~t = 1,\dots,L.
\end{equation}

Due to the existence of microstructure noise, we cannot observe $X_i(t)$ in reality. What we observe is  $Y_i(t_{ij})$, a noisy version of $X_i(\cdot)$ at times $t_{ij} \in (t-1, t ], j = 1,\dots ,n_i, i = 1,\dots ,d$. Here, $n_i$ is the total trading times and $t_{ij}$ is the $j$-th trading times of asset $i$ during a giving trading day $t$. The observations $Y_i(t_{ij})$ are allowed to be non-synchronized, i.e. $t_{i_1j}   \neq t_{i_2j}$ for any $i_1 \neq i_2$. In this article, we assume that
\begin{equation}
Y_i(t_{ij}) = X_i(t_{ij}) + \varepsilon_i(t_{ij}),
\end{equation}
where $\epsilon_i(t_{ij})$ are i.i.d. microstructure noise with mean zero and variance $\eta_i$.  $\varepsilon_i(\cdot)$ and $X_i(\cdot)$ are assumed to be independent with each other. With high-frequency data on the $t$-th day, we can calculate Threshold Averaging Realized Volatility Matrix (TARVM) estimator $\bm{\hat{\Sigma}}_y(t) $ as an estimator for $\bm{\Sigma}_x(t)$, which is shown to be consistent for the integrated RCOV matrix under certain conditions  \citep{tao2011large}.
To reduce the effective number of entries in $\bm{\Sigma}_x(t)$,  \citet{tao2011large} proposed a matrix-factor analysis (MFA) model as follows:
\begin{equation}
\label{equ:mfa}
\bm{\Sigma}_x(t) = \bm{A}\bm{\Sigma}_f(t)\bm{A}^{\top} + \bm{\Sigma}_0, ~~~~~t =1,\dots,L,
\end{equation}
where $\bm{\Sigma}_0$ is a $d\times d$ positive definite constant matrix, $\bm{\Sigma}_f(t)$ are $r\times r$ positive definite matrices and $\bm{A}$ is a $d\times r$ factor loading matrix. Here $r(<d)$ is usually a fixed small integer.  \eqref{equ:mfa} assumes that the daily dynamical structure of the matrix process $\bm{\Sigma}_x(t)$ is driven by that of a lower-dimensional latent process $\bm{\Sigma}_f(t)$, where $\bm{\Sigma}_0$ represents the static part of $\bm{\Sigma}_x(t)$.

Put
\begin{eqnarray}
\bar{\bm{\Sigma}}_x & = & \frac{1}{L}\sum_{t=1}^{L} \bm{\Sigma}_x(t),\\
\bar{\bm{S}}_x & = & \frac{1}{L}\sum_{t=1}^{L}\{  \bm{\Sigma}_x(t) -\bar{\bm{\Sigma}}_x \}^2,
\end{eqnarray}
and as $\bm{\Sigma}_x(t)$ is unknown in practice, we use $\hat{\bm{\Sigma}}_y(t)$ as a proxy. Let
\begin{eqnarray}
\bar{\bm{\Sigma}}_y & = & \frac{1}{L}\sum_{t=1}^{L} \hat{\bm{\Sigma}}_y(t),\\
\bar{\bm{S}}_y & = & \frac{1}{L}\sum_{t=1}^{L}\{  \hat{\bm{\Sigma}}_y(t) -\bar{\bm{\Sigma}}_y \}^2,\label{equ:sy}
\end{eqnarray}
where $\hat{\bm{\Sigma}}_y(t)$ is the RCOV matrix on day $t$ obtained by method TARVM from high-frequency data. Let $\hat{\bm{A}}$ be the $r$ orthonormal eigenvectors  of $\bar{\bm{S}}_y$, corresponding to the $r$ largest eigenvalues, as its columns. \citet{tao2011large} proposed the estimated factor volatilities as
\begin{equation}
\hat{\bm{\Sigma}}_f(t) = \hat{\bm{A}}^{\top}\hat{\bm{\Sigma}}_y(t) \hat{\bm{A}}, t=1,\dots,L,
\end{equation}
and the estimator for $ \bm{\Sigma}_0$ as
\begin{equation}
\hat{\bm{\Sigma}}_0 =  \bar{ \bm{\Sigma}}_y-  \hat{\bm{A}}\hat{\bm{A}}^{\top}\bar{\bm{\Sigma}}_y \hat{\bm{A}}\hat{\bm{A}}^{\top}.
\end{equation}

Then \citet{tao2011large} built up the dynamical structure of $\bm{\Sigma}_x$ by fitting a VAR model to the vectorized $\hat{\bm{\Sigma}}_f(t)$. For a $r\times r$ matrix $\bm{\Sigma}$, let $vech(\bm{\Sigma})$ be the $r(r+1)/2\times 1$ vector obtained by stacking together the lower triangle of $\bm{\Sigma}$ (including the diagonal). Then the VAR model for $\bm{\Sigma}_f(t)$ is of the form
\begin{equation}
\label{equ:var}
vech\{ \hat{\bm{\Sigma}}_f(t)  \} = \bm{\alpha}_0 + \sum_{j=1}^{q} \bm{\alpha}_j vech\{ \hat{\bm{\Sigma}}_f(t-j) \} + \bm{e}_t,
\end{equation}
where $q \ge 1$ is an integer, $\bm{\alpha}_0$ is a vector, $\bm{\alpha}_1, \dots, \bm{\alpha}_q$ are square matrices, and $\bm{e}_t$ is a white noise process with zero mean and finite fourth moments.

The VAR model cannot guarantee the positive definiteness of the forecasted RCOV matrix. To resolve this problem, \citet{KerenShen2020} proposed a
diagonal Conditional Autoregressive Wishart (DCAW) model to $\hat{\bm{\Sigma}}_f(t)$. Let $\mathcal{F}_{t -1 } = \sigma(\bm{\Sigma}_f(s), s < t)$ be the past history of the process at time $t$. Conditional on $\mathcal{F}_{t -1 }$, $\bm{\Sigma}_f(t)$ follows a central Wishart distribution
\begin{equation}
\bm{\Sigma}_f(t) | \mathcal{F}_{t -1 } \sim \mathcal{W}_n(\bm{\nu}, \bm{S}_f(t)/\bm{\nu}),
\label{equ:wischar}
\end{equation}
with $\bm{\nu}$ degrees of freedom and the scaling matrix $\bm{S}_f(t)$. Moreover, the scaling matrix
$\bm{S}_f(t)$ follows a linear BEKK recursion of order $(p,q)$:
\begin{equation}
\begin{split}
\label{equ:bekk}
\bm{S}_f(t) = &\bm{C} \bm{C}^{\top} + \sum_{i=1}^{p} \bm{B}_i\bm{S}_f(t-i)\bm{B}_i^{\top} \\
&+\sum_{j=1}^{q}\bm{A}_j\bm{\Sigma}_f(t-j)\bm{A}_j^{\top},
\end{split}
\end{equation}
where $\bm{A}_j, \bm{B}_i$ and $\bm{C}$ are all $r\times r$ matrices of coefficients. Different orders $(p,q)$ are used to make comparison among models.

The CAW process depends on the parameters $\{\bm{\nu},\bm{C},(\bm{B}_i)_{1\le i \le p},(\bm{A}_j)_{1\le j\le q}\}$, so that the total number of parameters is equal to $(p + q)r^2 + \frac{r(r+1)}{2} + 1 = \mathcal{O}(r^2)$ which still grows quick with the number of factors $r$ and the order $p$ and $q$. The DCAW model restricts the coefficient matrices $\bm{C},(\bm{B}_i)_{1\le i \le p},(\bm{A}_j)_{1\le j\le q}$ to diagonal matrices. Therefore, the number of parameters becomes $(p + q + 1)r + 1 = \mathcal{O}(r)$. The estimation of the parameters $\bm{\theta} = (\bm{\nu},diag(\bm{C})^{\top},diag(\bm{B}_i)^{\top}_{1\le i \le p},diag(\bm{A}_j)^{\top}_{1\le j\le q})^{\top}$ of the DCAW($p$,$q$) model is carried out by maximizing the log-likelihood function using the Broyden-Fletcher-Goldfarb-Shanno (BFGS) optimization procedure.

\subsection{CNN and ConvLSTM}
\label{sec:cnn}
Convolutional Neural Network (CNN) is a well known variety of deep neural network which has been widely used to extract local patterns from images.
The convolutional layer aims to learn feature representations of the inputs. It is composed of several convolution kernels which are used to compute different feature maps. Specifically, each neuron of a feature map is connected to a neighborhood of neurons in the previous layer.
The new feature map can be obtained by first convolving the input with a learned kernel and then applying an element-wise nonlinear activation function on the convolved results.
The activation function introduces nonlinearities to CNN, which are desirable for multi-layer networks to detect nonlinear features.

\citet{xingjian2015convolutional} extended the fully connected LSTM (FC-LSTM) to convolutional LSTM (ConvLSTM) which has convolutional structures in both the input-to-state and state-to-state transitions for precipitation nowcasting. Due to the inherent convolutional structure, the ConvLSTM layer is very suitable for spatiotemporal data. We follow the formulation of ConvLSTM as in \citet{xingjian2015convolutional}, which use $\mathcal{X}_{1}, ..., \mathcal{X}_{t}$ as inputs, $\mathcal{C}_{1}, ..., \mathcal{C}_{t}$ as cell outputs, $\mathcal{H}_{1}, ..., \mathcal{H}_{t}$ as hidden states. $i_t, f_t, o_t$ are input gate, forget gate and output gate respectively. Then the ConvLSTM is of the form:
\begin{equation}
\begin{aligned}
 & i_t = \sigma(W_{xi} \otimes \mathcal{X}_{t} + W_{hi} \otimes \mathcal{H}_{t-1} + W_{ci} \circ \mathcal{C}_{t-1} + b_i) \\
 & f_t = \sigma(W_{xf} \otimes \mathcal{X}_{t} + W_{hf} \otimes \mathcal{H}_{t-1} + W_{cf} \circ \mathcal{C}_{t-1} + b_f)  \\
 & \mathcal{C}_{t} = f_t \circ \mathcal{C}_{t-1} + i_t \circ tanh(W_{xc} \otimes \mathcal{X}_{t} + W_{hc} \otimes \mathcal{H}_{t-1} + b_c)  \\
 & o_t = \sigma(W_{xo} \otimes \mathcal{X}_{t} + W_{ho} \otimes \mathcal{H}_{t-1} + W_{co} \circ \mathcal{C}_{t} + b_o)  \\
 & \mathcal{H}_{t} = o_t \circ tanh(\mathcal{C}_{t})
\end{aligned}
\label{equ:convlstm}
\end{equation}
where $\otimes$ denotes the convolution operation, $\circ$ denotes the pointwise product. The ConvLSTM captures the local neighbors of the inputs and past states to determine the future state of a certain cell on a spatial grid as shown in (\ref{equ:convlstm}).

\section{Proposed Model}
\label{sec:ourmodel}
For simplicity, we use $\bm{\Sigma}_{t}$ to denote $\hat{\bm{\Sigma}}_y(t)$ which is viewed as a ground true RCOV matrix. Consider a stochastic, positive definite RCOV matrix $\bm{\Sigma}_{t}=(\Sigma_{{ij},t})$, with dimension $d \times d$ at time $t (t = 1,\dots,L)$. We now present our proposed model in the following aspects.

\subsection{Data Preprocessing}
The RCOV matrices are symmetric and positive definite, but the outputs of the end-to-end network are directly treated as the forecasting RCOV matrices, which can not guarantee the symmetry and positive definiteness. Therefore, before we train the network on the training set, data preprocessing methods must be applied to transform and compress the data.

To make sure the predicted RCOV matrices are positive definite, one method is to use Cholesky decomposition to transform the original RCOV matrices and then train the decomposed triangular matrices. The Cholesky decomposition for a $\mathbf{\Sigma}_{t}$ is of the form
\begin{equation}
\mathbf{\Sigma}_{t} =\bm{C}_{t} \bm{C}_{t}^{\top},
\end{equation}
where $\bm{C}_{t}$ is a lower triangular matrix with real and positive diagonal entries, and $\bm{C}_{t}^{\top}$ denotes the transpose of $\bm{C}_{t}$. After we get a predicted $\hat{\bm{C}_{t}}$, we could obtain a positive definite matrix $\hat{\bm{\Sigma}}_{t} = \hat{\bm{C}}_{t}\hat{\bm{C}}_{t}^{ \top}$. Note that Cholesky decomposition helps to decrease the range of the data. Such an operation reduces the large covariance values while boosts the small values, which makes the data suiting the neural networks better.

Recall that \citet{wang2016some} used the square root transformation to the RCOV matrix.
Consider the spectral decomposition of the RCOV matrix $\bm{\Sigma}_{t} = \bm{Q}_{t} \bm{\Lambda}_{t}\bm{Q}_{t}^{ \top}$, where $\bm{\Lambda}_{t} = diag(\lambda_{t,1}, \dots, \lambda_{t,p})$
is a diagonal matrix of the eigenvalues of $\bm{\Sigma}_{t}$, and $\bm{Q}_{t}$ is an orthonormal matrix consisting of eigenvectors of $\bm{\Sigma}_{t}$. Assume that $\lambda_{t,1} \ge \lambda_{t,2}  \ge \dots
\ge \lambda_{t,p} >0$. In our datasets, we observed the same phenomenon as in \citet{tao2013fast} that  the first several $\lambda_{t,i}$ are much bigger than the others. As the range is quite large, we found that taking square root of the eigenvalues could reduce the range. Denote $\bm{M}_{t}$ as $\bm{M}_{t}= diag(\lambda_{t,1}^{1/2}, \dots, \lambda_{t,p}^{1/2})$, and denote $\bm{O}_{t} = \bm{Q}_{t} \bm{M}_{t} \bm{Q}_{t}^{\top}$, then we have
\begin{equation}
\bm{O}_{t} = \bm{\Sigma}_{t}^{1/2}.
\end{equation}
Both Cholesky decomposition and square root transformation  could help maintain the positive definiteness of the predicted RCOV matrices and reduce  the range of data. These two transformations can be applied together, that is, we may apply square root transformation to the RCOV matrices and then apply Cholesky decomposition from the square root transformed matrices.
\subsection{Encoding-Generating Structure}
We name our proposed model Covariance Matrix Convolutional LSTM (CM-ConvLSTM) model. The proposed model has several appealing properties. First, its structure is intentionally designed with simplicity in mind, and yet provides superior accuracy compared with several traditional statistical methods.  Second, our model could handle higher dimensional RCOV matrix but traditional methods could not. Our model could deal with high-dimensional and low-dimensional RCOV matrices in a consistent way. Third, with moderate numbers of filters and layers, our method achieves remarkably fast speed for practical on-line usage. Our model is easy to implement with the excellent open deep learning packages.

Given past $m$ observations, a 3D matrix $\tilde{\bm{\Sigma}}_{t} = [\bm{\Sigma}_{t}; \bm{\Sigma}_{t-1};\dots;\bm{\Sigma}_{t-m+1}]$ with dimension $m\times d \times d$, the corresponding transformed 3D matrix $\tilde{\bm{L}}_{t} = [\bm{L}_{t}; \bm{L}_{t-1};\dots;\bm{L}_{t-m+1}]$ can be obtained using the data preprocessing methods mentioned above. The input of the network is the transformed matrix $\tilde{\bm{L}}_{t}$, and our model is an end-to-end model. Therefore, the output of the network is $\bm{L}_{t+1}$. It is easy to obtain the target prediction $\bm{\Sigma}_{t+1}$ after applying the same transformations in the data preprocessing stage in the reverse order. We wish to learn a mapping $F$, which conceptually consists of the following operations.

\subsubsection{Encoding}
The network consists of two types of encoding layers: ConvLSTM encoding layer and CNN encoding layer. We first use a ConvLSTM layer to operate on the inputs which can extract the local information of the inputs and past states at each timestamp. This layer aims to learn the complex spatiotemporal patterns.

Then we apply one or more CNN layers which work as the local structure representation and non-linear mapping layer which is equivalent to convolving the input feature maps by a set of filters. These layers aim to further capture the local information and conduct non-linear mapping on the features learnt in the ConvLSTM layer.
We apply the Leaky Rectified Linear Unit (LReLU) as the activation function on the filter responses after each ConvLSTM layer or CNN layer.
Zero-padding is applied to make sure the output feature maps are the same size as the input transformed RCOV matrices.
It is possible to add more CNN layers to increase the non-linearity according to the complexity of the data. But this can increase the complexity of the model, and thus demands more training time.

\subsubsection{Generating}
In the last stage, the predicted overlapping broader local patches are averaged to produce the final transformed RCOV matrix. The averaging can be considered as a pre-defined filter on a set of feature maps.
Motivated by this, we define a convolutional layer to generate the transformed RCOV matrix $\bm{L}_{t+1}$ and so that obtain the final prediction RCOV matrix $\bm{\Sigma}_{t+1}$.
In this layer, instead of using non-linear LReLU activation function, we use linear mapping here to act like regression on the learned feature maps in the previous layer. By stacking multiple ConvLSTM layers and CNN layers, we can build the end-to-end network with the encoding-generating structure.

\begin{figure}[ht]
\vskip 0.2in
\begin{center}
\centerline{\includegraphics[width=18cm]{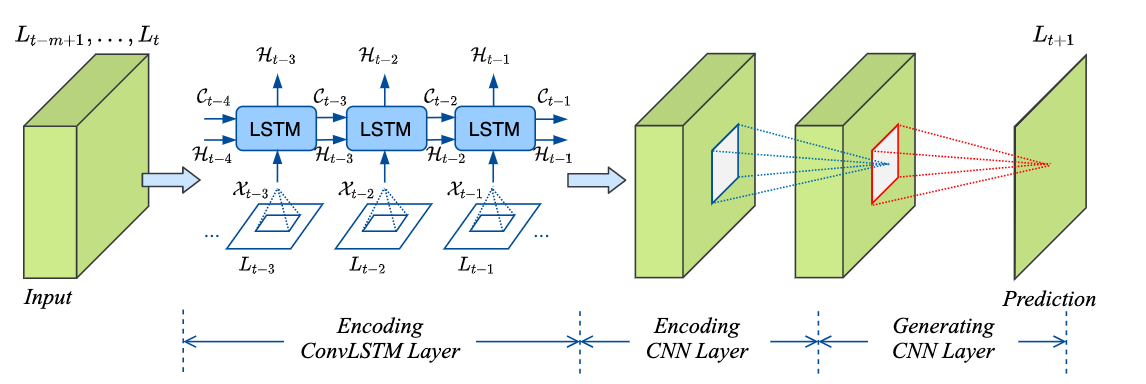}}
\caption{Diagram of the proposed encoding-generating model for RCOV matrices forecasting.}
\label{fig:the_model}
\end{center}
\vskip -0.2in
\end{figure}

An overview of the network is depicted in Figure \ref{fig:the_model}. This shows how the above operations form a ConvLSTM and CNN based network. Unlike other CNN architectures, our model has no pooling layers or fully-connected layers. Zero padding is needed before a convolution operation in the model to ensure that the ouputs of the network have the same size as the inputs. 

\subsection{Training}
\label{sec:training}
Learning the end-to-end mapping function $F$ requires the estimation of network parameters.
This is achieved through minimizing the loss between the predicted and the corresponding ground truth matrices.
It is worth noting that deep learning models do not preclude the usage of other kinds of loss functions, if only the loss functions are derivable. If a better perceptually motivated metric is given during training, it is flexible for the network to adapt to that metric.

\begin{figure}[ht]
\vskip 0.2in
\begin{center}
\centerline{\includegraphics[width=8cm]{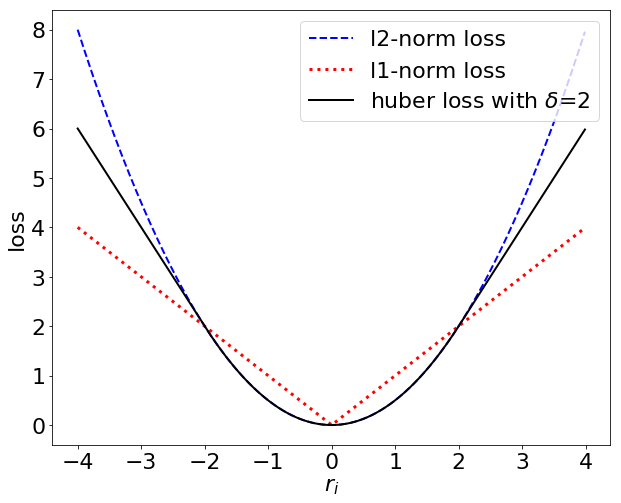}}
\caption{Huber loss (black solid line), L1-norm loss (red dotted line) and L2-norm loss (blue dashed line) as a function of $r_i$.}
\label{fig:lossfunctions}
\end{center}
\vskip -0.2in
\end{figure}

We were troubled by ``outliers'' in the RCOV matrices even after data preprocessing. The network would tend to struggle between the majority of small values and small proportion of extreme big values that the performances stuck and are hard to improve. A popular robust loss function called Huber Loss is chosen instead of L1-norm loss and L2-norm loss to relieve this situation.  Huber Loss is resistant to outliers in the data, while it maintains sensitive to the small values. And this sensitivity  enables this loss function to perform well on our real-world datasets.
Mathematically, the Huber loss function is defined as
\begin{equation}
\ell_{\delta}(r_i) =
\left\{
 \begin{array}{ll}\frac{1}{2}(r_i)^2,&|r_i|\leq \delta \\
 \\
\delta(|r_i|-\frac{1}{2}\delta),& otherwise
 \end{array}
  \right.
\label{equ:huber}
\end{equation}
where $r_i$ is the error of the prediction of the $i$-th observation.
Figure \ref{fig:lossfunctions} distinguishes Huber loss from L1-norm loss and L2-norm loss in terms of $r_i$.
When the absolute difference between the observed and predicted values is larger than a prespecified value $\delta$, the Huber loss is reduced to the L1-norm loss in order to prevent the squared differences with large magnitude which may dominate the whole loss.
As our data is a time series of matrices, the form of Huber loss function in (\ref{equ:huber}) has to be adapted to the 2D observed and predicted matrices. Given a set of pairs of $\{\bm{\hat{Y_i}}, \bm{Y_i} \}$, where $\bm{\hat{Y_i}}$ and $\bm{Y_i}$ are the predicted matrix and  the corresponding ground true matrix of the $i^{th}$ observation, the Huber loss function becomes:
\begin{equation}
\ell_t(\Theta) =
\left\{
 \begin{array}{ll}\frac{1}{2}||\bm{\hat{Y_i}} - \bm{Y_i}||^2,&||\bm{\hat{Y_i}} - \bm{Y_i}||\leq \delta \\
 \\
\delta(||\bm{\hat{Y_i}} - \bm{Y_i}||-\frac{1}{2}\delta),& otherwise
 \end{array}
  \right.
  \label{equ:huber_mx}
\end{equation}
where $||\bm{\hat{Y_i}} - \bm{Y_i}||$ and $||\bm{\hat{Y_i}} - \bm{Y_i}||^2$ are the L1-norm loss and L2-norm loss of the cells in these two matrices, respectively.
Gradient descent method based on Adam optimization algorithm and L1 regularization are used in the training process.

\section{Experimental Results}
\label{sec:emprical}
In this section, we study the forecasting performances of our model on synthetic data and real data. The root mean squared error (RMSE) and mean absolute error (MAE) are adopted as the evaluation metrics. Given a pair of predicted and ground true matrices, the RMSE takes square root after the sum of the squared errors of the corresponding cells, while the MAE is the sum of the absolute errors.

\subsection{Experiments on Synthetic Data} 
\label{subsec:simulations}
\subsubsection{Simulations}
To explore the superiority of our model, we first conduct simulation studies in this section. Firstly, we generate a dataset based on the Wishart distribution in  \eqref{equ:wischar} and DCAW model  in \eqref{equ:bekk}. Secondly, we change the underlying Wishart distribution to matrix-F distribtion \citep{zhou2019time} and generate another dataset. That is,
\begin{equation}
\bm{\Sigma}_f(t) | \mathcal{F}_{t -1 } \sim F(\bm{\nu}_1, \bm{\nu}_2, \frac{\bm{\nu}_2-n-1}{\bm{\nu}_1}\bm{S}_f(t)),
\label{equ:f}
\end{equation}
with $E(\bm{\Sigma}_f(t) | \mathcal{F}_{t -1 } = \bm{S}_f(t))$ and  $\bm{S}_f(t)$ is defined in  \eqref{equ:bekk}. The conditional distribution of $\hat{\bm{\Sigma}}_f(t)$ is matrix-F with a  BEKK mean structure. Following \citet{konno1991note} and \citet{leung1996identity}, $\bm{\Sigma}_f(t)$ in the above equation could be written as
\begin{equation}
\begin{split}
\bm{\Sigma}_f(t) = & \left( \frac{\bm{\nu}_2-n-1}{\bm{\nu}_1} \right) \times \bm{S}_f(t)^{1/2}L(t)^{1/2}R(t)^{-1}  \\
& \times L(t)^{1/2}\bm{S}_f(t)^{1/2}
\label{equ:recusionf}
\end{split}
\end{equation}
where $L(t) \sim Wishart(\bm{\nu}_1, \bm{I}_n)$, $R(t) \sim Wishart(\bm{\nu}_2, \bm{I}_n)$ are independent and  $\bm{I}_n$ is the $n\times n$ identity matrix.

In both sets of simulations, the loading matrix $\bm{A}$ is formed by eigenvector corresponding to the largest eigenvalues of $\bar{\bm{S}}_y$  in \eqref{equ:sy} obtained from the S\&P100 dataset (see Section \ref{sec:real_data}). We set $r=3$ and $(p, q)$ as $(2, 2)$, while the matrices of coefficients are set as
\[
\bm{A}_1 =
\begin{bmatrix}
0.2  & 0    & 0\\
0    & 0.4  & 0\\
0    & 0    & 0.5
\end{bmatrix} ,
\bm{A}_2 =
\begin{bmatrix}
0.3  & 0     & 0\\
0    & 0.5   & 0\\
0    & 0     & 0.2
\end{bmatrix}
\]
\[
\bm{B}_1 =
\begin{bmatrix}
0.2  & 0     & 0\\
0    & 0.5   & 0\\
0    & 0     & 0.4
\end{bmatrix} ,
\bm{B}_2 =
\begin{bmatrix}
0.3  & 0     & 0\\
0    & 0.5   & 0\\
0    & 0     & 0.2
\end{bmatrix}
\]
\[
\bm{C}=
\begin{bmatrix}
0.5    & 0.2     & 0.3\\
0.2    & 0.5     & 0.25\\
0.3    & 0.25    & 0.5
\end{bmatrix}
\]

The initial values of $\bm{S}_f(t)$ are generated by the first two days' data in S\&P100 dataset. In the first simulation study, we set $\bm{\nu}=5$. In the second simulation study, we set $\bm{\nu}_1=10, \bm{\nu}_2=8$. Each simulation has $L=5000$ samples. The simulation from Wishart distribution can be discribed in the following steps:

\begin{enumerate}
	\item[1.] Assign values to parameters,
	\item[2.] For $t=1$, set the initial value to $\bm{S}_f(0)$ and $\bm{S}_f(1)$, sample $\bm{\Sigma}_f(0)$ and $\bm{\Sigma}_f(1)$ by  \eqref{equ:wischar}; otherwise sample $\bm{\Sigma}_f(t-1)$ based on $\bm{S}_f(t-1)$,
	\item[3.] Update  $\bm{S}_f(t)$  from the BEKK recursion    in \eqref{equ:bekk},
	\item[4.] Set $t=t+1$ and repeat steps 2 and 3.
\end{enumerate}
The simulation from F distribution would change step 2 by sampling $L(t)$ and $R(t)$ from two independent Wishart distributions first and then obtain $\bm{\Sigma}_f(t)$ by \eqref{equ:recusionf}.

\subsubsection{Implementation Details and Results}
We conduct 10 simulations for each set of distribution and split 70\%, 10\% and 20\% of the 5000 samples as training, validation and testing sets, respectively.  The parameters of networks are optimized by Adam optimization algorithm with weight decay of 1e-5, $\beta_1$ of 0.9, $\beta_2$ of 0.999, the initial learning rate of 0.001 and mini-batch sizes of 128. The $\lambda$ of L1 regularization is set as 0.005 and the $\delta$ of huber loss is set as 300. As all the samples are drawn from distributions, the simulated data do not include "outliers". In this case, the $\delta$ of huber loss tends to be larger (L2-norm loss). Table \ref{tab:simu_struc} illustrates the standard three-layer structure of CM-ConvLSTM used in the simulations.

The evaluations in terms of RMSE of these 10 simulated data are displayed in Table \ref{tab:simu}. Whether under Wishart distribution samples or F distribution samples, our proposed CM-ConvLSTM model consistently outperforms DCAW model, by about 0.85 and 0.66 in terms of average RMSE over these two distribution samples, respectively. These observations support our expectation that our model has less dependence on the underlying distribution and is more robust. 
In the next section, the superiority of our model will be further demonstrated on three real-world datasets.
In fact, the real data is too complex to be modeled by a simple Wishart distribution.  Our model, however, makes no distributional assumption.

\begin{table}
	\caption{\label{tab:simu_struc}ConvLSTM architecture for simulations with lag length 20. }
\vskip 0.15in
\begin{center}
\scriptsize
	\begin{tabular}{clllcc}
		\toprule
		{\bf Layer} 	&	{\bf Operation} & {\bf Size-in}& {\bf Size-out}& {\bf Kernel size} & {\bf Number of kernels}\\
		\midrule
		\multirow{2}{*}{1}  &  ConvLSTM  &  $20\times1 \times60\times60$  &  $8\times60\times60$  &  $ 3\times3$  &  8  \\	
		&  leaky-relu  &  $8\times60\times60$  &  $8\times60\times60$  & - & -  \\
		\multirow{2}{*}{2}  &  Conv  & $8\times60\times60$  &  $16\times60\times60$  &  $3\times3$ &  16  \\	
		&  leaky-relu  &  $16\times60\times60$  &  $16\times60\times60$  & - & -  \\
		\multirow{2}{*}{3}  &  Conv  & $16\times60\times60$  &  $1\times60\times60$  &  $1\times1$ &  1  \\	
		&  linear  &  $1\times60\times60$  &  $1\times60\times60$  & - & -  \\
		\bottomrule
	\end{tabular}
\end{center}
\vskip -0.1in
\end{table}

\begin{table*}[t]
	\caption{Performance comparison of average RMSE for two set of simulation studies, where W and F denote Wishart distribution and matrix-F distribtion respectively.}
	\label{tab:simu}
\vskip 0.15in
\begin{center}
\begin{small}
	\begin{tabular}{lccccccccccc}
		\toprule
		&  1  &  2 &  3  & 4  & 5  & 6 & 7  & 8  & 9  & 10  \\
		\midrule
		DCAW (W) & 1.102 & 1.635 & 1.101 & 1.148 &	1.096 & 1.083 & 1.452 & 1.062 & 1.265 & 1.274  \\
		CM-ConvLSTM (W) & 0.359 & 0.360 & 0.362 & 0.395 & 0.371 & 0.335 & 0.351 & 0.374 & 0.357 & 0.411  \\
		difference (W) & 0.743 & 1.275 & 0.739 & 0.753 & 0.725 & 0.748 & 1.101 & 0.688 & 0.908 & 0.863  \\ 
		\midrule
		DCAW (F) & 1.473 & 1.350 & 1.575 & 1.534 & 1.559 & 1.770 & 1.602 & 1.887 & 1.620 & 1.595  \\
		CM-ConvLSTM (F) & 0.877 & 0.709 & 0.881 & 0.841 & 0.930 & 1.150 & 0.779 & 1.164 & 1.041 & 1.018  \\
		difference (F) & 0.596 & 0.642 & 0.694 & 0.693 & 0.629 & 0.621 & 0.822 & 0.723 & 0.579 & 0.577  \\
\bottomrule
\end{tabular} 
\end{small}
\end{center}
\vskip -0.1in
\end{table*}

\subsection{Experiments on Real Data}
We further show the outstanding forecasting performances of our model by comparing the performances of some traditional time series models and some well-known deep learning models on three real-world datasets. All the deep learning models are trained on GPU Tesla V100 SXM2 with about 32GB memory.

\subsubsection{Data Description}
\label{sec:real_data}
We consider three datasets:  (i) all constituent stocks of the DJIA index, (ii) all constituent stocks of the S\&P100 index, (iii) all constituent stocks of the S\&P500 index.
All three datasets are downloaded from the NYSE TAQ database of WRDS, and daily RCOV matrices are calculated using the ARVM method \citep{wang2010vast}. The trading records are from 9:30 am to 4:00 pm each day, with observations before 10:00 am deleted to avoid opening effects. The sampling frequency is set to five minutes. Stocks with less than 100 daily trading records are also deleted.

The first dataset comprises the 30 constituent stocks of the DJIA index from 2007 to the end of 2013. Only 25 stocks have full intra-day price data from January 18, 2007 to December 31, 2013 (1752 trading days).
RCOV matrices ($\bm{\Sigma}_{t}, t = 1, \dots , 1752$) with dimension equals to 25 are obtained.
The second dataset contains intra-day data on 60 constituent stocks ($d = 60$) of the S\&P100 index, which have full records during the period from September 10, 2003 to December 30, 2016($L = 3345$).
The third dataset contains a selection of S\&P500 constituent stocks that continuously traded over the period from September 10, 2003 to December 30, 2016. The full dataset contains 244 stocks ($d = 244$) and 3345 observations ($L = 3345$). 

\begin{figure}[ht]
\vskip 0.2in
\begin{center}
\centerline{\includegraphics[width=10cm]{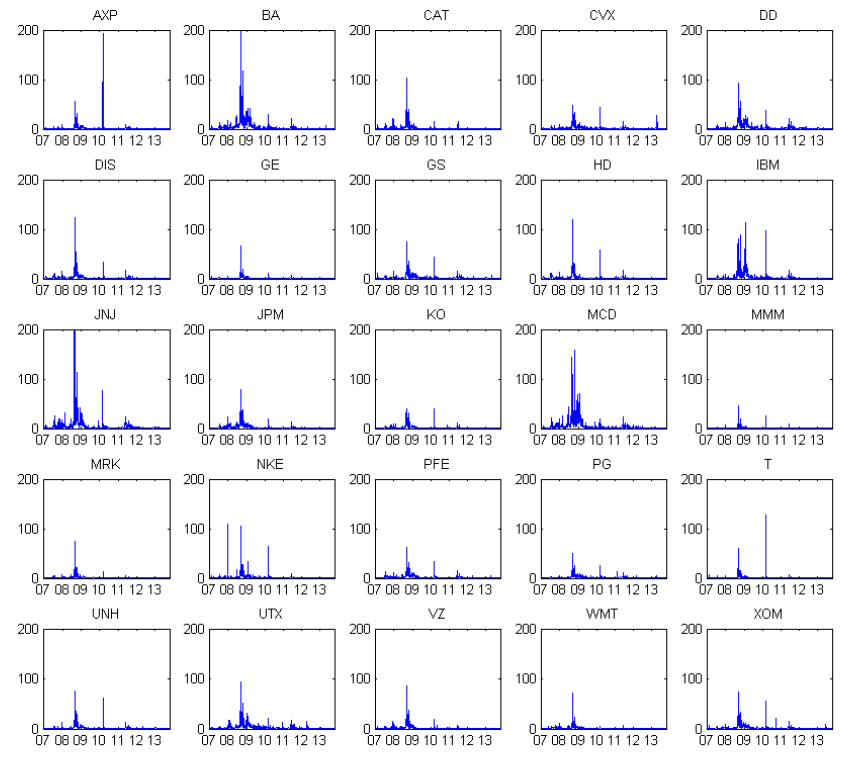}}
\caption{Realized variances of the DJIA constituents from 01/18/2007 - 12/31/2013}
\label{fig:stocks_volatility}
\end{center}
\vskip -0.2in
\end{figure}

The realized variances vary a lot, in some days, the variances (diagonal values) of some stocks jump to another magnitude  which distract the neural network learning.  Figure \ref{fig:stocks_volatility} reveals high volatilities around the 2008 subprime mortgage crisis and during the flash crash on May 6,2010 for DJIA dataset. The same phenomenon exists in the other two datasets.
The dimensions of the three datasets increase from 25 to 60, and finally 244, aimed to demonstrate the power of our model in handling high dimensional RCOV matrices. All datasets are split into three parts: the testing set contains the last $252$ days' RCOV matrices, the validation set contains the second last $252$ days' RCOV matrices, the remaining matrices all go to the training set.

\subsubsection{Models for Comparison}
For the sake of model comparison, besides the proposed model, we also have four time series models and three CNN-based deep learning models.
The time series models are moving average (MA), exponential moving average (EMA), MFA-VAR and MFA-DCAW. For deep learning models, some well-known networks, including CNN, ResNet and DenseNet are considered to apply to the datasets.

\subsubsection{Implementation Details and Ablation Study}
In the following, we conduct ablation experiments to gain a better understanding of the impact of using different data preprocessing methods and loss functions. Due to the limited time and computing resource, we only conduct the comprehensive comparisons on  the DJIA dataset instead of all three datasets.
The preprocessing methods include no preprocessing, Choleskey decomposition, square root transformation  and  doing  Choleskey  decomposition  after  square  root  transformation. The  loss functions  include  L1-norm  loss,  L2-norm  loss,  and  Huber  loss.
We systematically tested different network architectures of different combinations of filter  size,  the  number  of  filters  and the  number  of  layers on DJIA dataset.
For S\&P100 and S\&P500 datasets, less combinations are tested based on the experience of DJIA dataset.

Tables \ref{tab:preprocess_DJIA} and \ref{tab:lossfunctions_DJIA} show the average RMSE and MAE for each preprocessing method and loss function on DJIA dataset, respectively. We can find that using Cholesky decomposition after square root transformation of eigenvalues gives the best performance. The Huber loss performs the best among all the loss functions. Similar findings can be observed on S\&P100 and S\&P500 datasets.


\begin{table}[t]
	\caption{Evaluation of the impact of data preprocessing methods on DJIA validation set.}
	\label{tab:preprocess_DJIA}
\vskip 0.15in
\begin{center}
\begin{small}
\begin{tabular}{lcc}
\toprule
Preprocessing  &  RMSE  &  MAE  \\
\midrule
$\bm{\Sigma}_{t}$  &  3.974  &  76.885  \\
$\bm{\Sigma}_{t} \rightarrow \bm{L^1_{t}} \mathtt{(Cholesky~Decomposition)}$  &  3.774  &  72.613  \\
$\bm{\Sigma}_{t} \rightarrow \bm{L^2_{t}} \mathtt{(Square~ Root)}$  &  3.792  &  72.709  \\
$\bm{\Sigma}_{t} \rightarrow \bm{L^2_{t}} \rightarrow \bm{L^1_{t}}$  &  \textbf{3.760}  &  \textbf{72.139}  \\
\bottomrule
\end{tabular}
\end{small}
\end{center}
\vskip -0.1in
\end{table}
\begin{table}[!htbp]
    \caption{Evaluation of the impact of loss functions on DJIA validation set.} 
    \label{tab:lossfunctions_DJIA}
\vskip 0.15in
\begin{center}
\begin{small}
\begin{tabular}{lcc}
\toprule
Loss Function  &  RMSE  &  MAE  \\
\midrule
L1-norm loss  &  3.799  &  72.812  \\
L2-norm loss  &  3.896  &  75.437  \\
Huber loss  &  \textbf{3.780}  &  \textbf{72.512}  \\
\bottomrule
\end{tabular}
\end{small}
\end{center}
\vskip -0.1in
\end{table}
The parameters of networks are optimized by Adam optimization algorithm with weight decay of 1e-5, $\beta_1$ of 0.9 and $\beta_2$ of 0.999. The initial learning rates are 0.001, 5e-4, 5e-4 and mini-batch sizes are 128, 128, 64 for DJIA, S\&P100 and S\&P500, respectively. We use L1 regularization in the training process and we set the $\lambda$ as 0.005, 0.001, 0.0005 and the $\delta$ of huber loss as 800, 2400, 12000 for these three datasets respectively. 

The architecture of the ConvLSTM model is very simple and is determined by the validation RMSE and MAE. For DJIA dataset, the configuration is a standard three-layer network, the same as in Figure \ref{fig:the_model}, with the input lag length of 20, as shown in Table \ref{tab:djia_struc}. We find that adding more CNN encoding layers can not further improve the model performance, then we stop at three layers for this dataset.
For S\&P100 and S\&P500 datasets, the networks have 1 or 2 more CNN encoding layers respectively, and the architecture of S\&P100 is explicated in Table \ref{tab:sp100_struc}. The state-to-state kernels in the ConvLSTM encoding layer are of size $5 \times 5$ with larger receptive field, showing the spatiotemporal correlations in the RCOVs.
As time advances, the later states have the much larger reception fields.

\begin{table}
	\caption{\label{tab:djia_struc}ConvLSTM architecture for DJIA with lag length 20.  ``Conv'' refers the layer of convolution. }
\vskip 0.15in
\begin{center}
\scriptsize
	\begin{tabular}{clllcc}
		\toprule
		{\bf Layer} 	&	{\bf Operation} & {\bf Size-in}& {\bf Size-out}& {\bf Kernel size} & {\bf Number of kernels}\\
		\midrule
		\multirow{2}{*}{1}  &  ConvLSTM  &  $20\times1\times25\times25$  &  $4\times 25\times25$  &  $3\times3$  &  4  \\	
		&  leaky-relu  &  $4\times25\times25$  &  $4 \times25\times25$  & - & -  \\
		\multirow{2}{*}{2}  &  Conv  & $4\times25\times25$  &  $8\times25\times25$  &  $3\times3$ &  8  \\	
		&  leaky-relu  &  $8\times25\times25$  &  $8\times25\times25$  & - & -  \\
		\multirow{2}{*}{3}  &  Conv  &  $8\times25\times 25$  &  $1\times25\times25$  &  $1\times1$  &  1  \\	
		&  linear  &  $1\times25\times25$  &  $1\times25\times25$  & - & -  \\
		\bottomrule
	\end{tabular}
\end{center}
\vskip -0.1in
\end{table}

\begin{table}
	\caption{\label{tab:sp100_struc}ConvLSTM architecture for S\&P100 with lag length 20. }
\vskip 0.15in
\begin{center}
\scriptsize
	\begin{tabular}{clllcc}
		\toprule
		{\bf Layer} 	&	{\bf Operation} & {\bf Size-in}& {\bf Size-out}& {\bf Kernel size} & {\bf Number of kernels}\\
		\midrule
		\multirow{2}{*}{1}  &  ConvLSTM  &  $20\times1 \times60\times60$  &  $16\times60\times60$  &  $ 5\times5$  &  16  \\	
		&  leaky-relu  &  $16\times60\times60$  &  $16\times60\times60$  & - & -  \\
		\multirow{2}{*}{2}  &  Conv  & $16\times60\times60$  &  $16\times60\times60$  &  $3\times3$ &  16  \\	
		&  leaky-relu  &  $16\times60\times60$  &  $16\times60\times60$  & - & -  \\
		\multirow{2}{*}{3}  &  Conv  & $16\times60\times60$  &  $32\times60\times60$  &  $3\times3$ &  32  \\	
		&  leaky-relu  &  $32\times60\times60$  &  $32\times60\times60$  & - & -  \\
		\multirow{2}{*}{4}  &  Conv  & $32\times60\times60$  &  $1\times60\times60$  &  $5\times5$ &  1  \\	
		&  linear  &  $1\times60\times60$  &  $1\times60\times60$  & - & -  \\
		\bottomrule
	\end{tabular}
\end{center}
\vskip -0.1in
\end{table}

Three well-known networks including CNN, ResNet and DenseNet are used to compare with our proposed model. We first illustrate the structure of CNN model. In our experiments, we find that the kernel size in the first layer tends to be larger ($7 \times 7$ or $9 \times 9$), verifying that the existence of local structure of RCOVs have positive effects on forecasting performances. While the following layers are mainly non-linear mapping of the representation of the features learnt before with $1 \times 1$ or $3 \times 3$ kernel size. L1 regularization is used to determine the number of kernels. For DJIA, the model only has two layers and the numbers of kernels in the first and second layer are 2 and 1, respectively. For S\&P100 and S\&P500 datasets, the architectures are very similar, which have four and five layers, respectively. Based on the structure of CNN, short connections are added to build the corresponding ResNet and DenseNet model. 
To avoid the notable increase of the network parameters of DenseNet for S\&P100 and S\&P500 datasets, we employ 1 kernel with size $1\times1$ to the feature maps before they are connected to each layer.

\subsubsection{Forecasting Performances}
Table \ref{tab:compare_models} shows the RMSEs of all the mentioned models on DJIA, S\&P100 and S\&P500 testing sets.
The best values of lag length are listed in the brackets next to the RMSEs of MA and EMA models in the table, which are selected according to the performances on validation sets. The values listed in the brackets of the MFA-VAR models are the orders $q$ in \eqref{equ:var} and the dimension of factor matrix in \eqref{equ:mfa}. For example, (3, 1) means VAR(3) based on $r=1$ factor matrix is chosen by the BIC criterion on training sets recommended by \citet{KerenShen2020}. Similarly, for MFA-DCAW model, (1, 1, 1) means the order $(p, q)$ of DCAW in \eqref{equ:bekk} is (1, 1) based on $r=1$ factor matrix, which is chosen by the BIC criterion on training sets. The architectures of the deep learning models for the three datasets are selected with the smallest RMSE on validation sets, and the number of parameters is listed next to the RMSE.

\begin{table*}[t]
\caption{Forecasting performances of different models on DJIA, S\&P100 and S\&P500 testing sets.}
\label{tab:compare_models}
\vskip 0.15in
\begin{center}
\begin{small}
\resizebox{\textwidth}{26mm}{
\begin{tabular}{cl|cc|cc|cc}
\toprule
\multicolumn{2}{c|}{Dataset}  & \multicolumn{2}{c|}{DJIA}  & \multicolumn{2}{c|}{S\&P100}  & \multicolumn{2}{c}{S\&P500}   \\
\multicolumn{2}{c|}{Model}                                 & \multicolumn{1}{c}{Parameters} & \multicolumn{1}{c|}{RMSE} & \multicolumn{1}{c}{Parameters} & \multicolumn{1}{c|}{RMSE} & \multicolumn{1}{c}{Parameters} & \multicolumn{1}{c}{RMSE} \\
\midrule
\multirow{4}{*}{\begin{tabular}[c]{@{}c@{}}Time Series \\ Model\end{tabular}}
&  MA  &  (7)  &  3.851  &  (3)  &  17.430  &  (5)  &  110.689
\\
&  EMA  &  (7)  &  3.734  &  (5)  &  16.277  &  (6)  &  107.586
\\
&  MFA-VAR  &  (3, 1)  &  9.542  &  (3, 1)  &  26.173  &  (3, 1)  &  136.627
\\
&  MFA-DCAW  &  (1, 1, 1)  &  7.758  &  (1, 1, 1)  &  25.852  &  (1, 1, 1)  &  134.200
\\
\midrule
\multirow{4}{*}{\begin{tabular}[c]{@{}c@{}}Deep Learning \\ Model\end{tabular}}
&  CNN  &  4052  &  3.397  &  41376  &  15.054  &  61456  &  105.186
\\
&  ResNet  &  4078  &  3.407  &  41712  &  15.577  &  62032  &  106.593
\\
&  DenseNet  &  4329  &  3.401  &  42276  &  15.073  &  62526  &  106.735
\\
&   \textbf{ConvLSTM}  &   \textbf{1016}  &  \textbf{3.394}  &   \textbf{34912}  &  \textbf{15.016}  &   \textbf{44128}  &  \textbf{104.924}
\\
\bottomrule
\end{tabular}
}
\end{small}
\end{center}
\vskip -0.1in
\end{table*}

\begin{figure}[ht]
\vskip 0.2in
\begin{center}
    \centerline{\includegraphics[width=16cm]{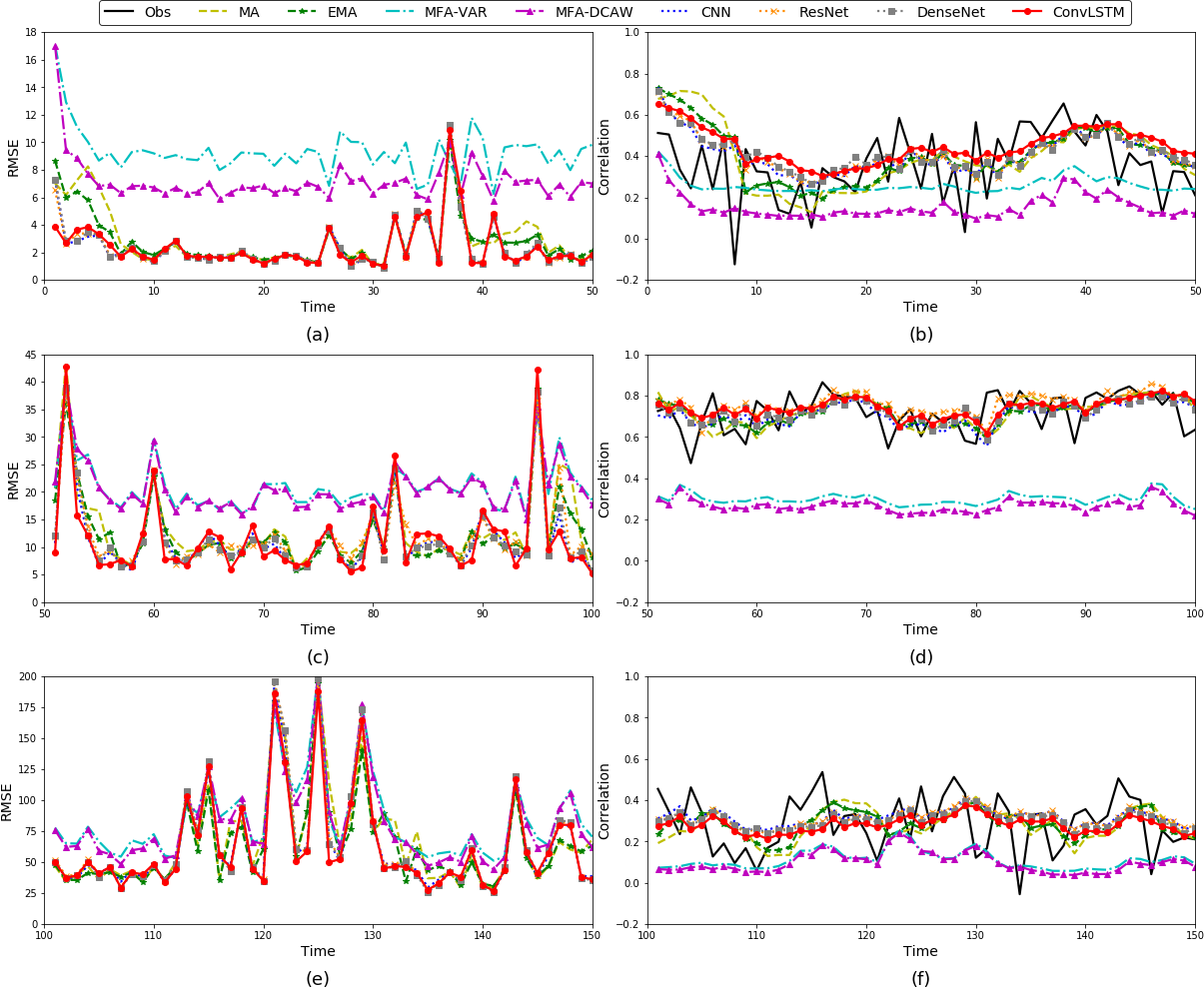}}
    \caption{Comparison of different models
    based on RMSE and correlation over time on DJIA ((a) and (b)), S\&P100 ((c) and (d)) and S\&P500 datasets ((e) and (f))(Best viewed in color).}
    \label{fig:all_daily_error}
\end{center}
\vskip -0.2in
\end{figure}

Among these models, our proposed model achieved the best RMSE on all three datasets. Our experiments show that the deep learning models perform consistently better than the time series models, and our ConvLSTM model has the minimum number of network parameters in these four end-to-end deep learning models. 
Figure \ref{fig:all_daily_error} compares different models by using RMSE and correlation over time. Here, we show differnt part of time of the 252-day testing data.
According to the daily RMSE in the 252-day testing data, our model performs the best among these days on all three datasets except for some days with extremely large volatilities. These ``outliers'' come from some positive efforts from the US Government to stimulate the economy after the crisis. We randomly select two stocks for each dataset, which are AXP and BA for DJIA, ABT and ACN for S\&P100, AVP and HAL for S\&P500, and then calculate the correlation over time. From Figure \ref{fig:all_daily_error}, we can observe that our ConvLSTM model catch the trend of the correlation over time.

\begin{figure}[ht]
\vskip 0.2in
\begin{center}
    \centerline{\includegraphics[width=8cm]{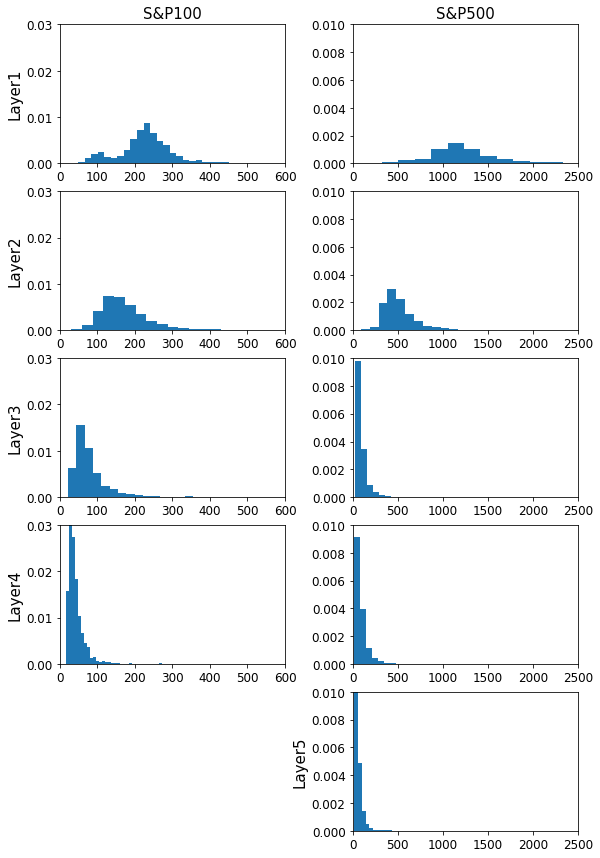}}
    \caption{The histograms of distances of feature maps and  ground true RCOV matrices over four and five layers for S\&P100 dataset and  S\&P500 dataset, respectively.}
    \label{fig:sp100500_distance}
\end{center}
\vskip -0.2in
\end{figure}

For S\&P100 and S\&P500 datasets,
the histograms of the distances of feature maps and observed RCOV matrices of each layer are shown in Figure \ref{fig:sp100500_distance}, where the distance of each feature map and observed RCOV matrix is defined by the MAE. Each histogram is ranged in the same scale for comparison. We find that the distances are smaller layer by layer, and the output from the last layer shows the smallest distance. Although the distances of the third and fourth layer of S\&P500 dataset look similar, but we find that the distances of 60\% of the covariances in the RCOV matrix are actually smaller after the third layer.

\section{Conclusion and Future Work}
\label{sec:conclusion}
In this paper, we have successfully built an end-to-end deep learning model to tackle the RCOV matrices forecasting problem. To our knowledge, our study is the first attempt to apply ConvLSTM and CNN to model the RCOV matrices. The proposed model could fit RCOV matrices well and empirically show its outstanding performances on simulations and three real-world datasets of low to high dimensions.

For future work,
the best method of ordering the stocks with a reasonable complexity needs further exploration.
In this paper, several methods were used to order the stocks in the RCOV matrix. The first method is to sort the stocks by their average daily variances. The second method is to take the first PCA component of the average RCOV matrices and use its coefficients to sort the stocks. The third method is to use a block modeling method in social network clustering that tries to find an ordering of stocks so that the clusters of stocks become apparent as 'blocks'.
We tested all the proposed methods of ordering the stocks and some random orders.
Based on the results, we found that changing the ordering of stocks would affect the performances slightly ($\pm 0.4 $ in MAE). This may be explained by the market globalization and most of the stocks are weakly correlated (correlations ranged from 0.27 to 0.59) and using filters on a local block of RCOV matrices could be sufficient in modeling and forecasting RCOV matrices.

\newpage
{\small
\bibliography{reference}
\bibliographystyle{plainnat}}

\end{document}